\documentclass[pdflatex,sn-mathphys-num, iicol]{sn-jnl}% Math and Physical Sciences Numbered Reference Style 
\usepackage{graphicx}%
\usepackage{multirow}%
\usepackage{amsmath,amssymb,amsfonts}%
\usepackage{amsthm}%
\usepackage{mathrsfs}%
\usepackage[title]{appendix}%
\usepackage{xcolor}%
\usepackage{textcomp}%
\usepackage{manyfoot}%
\usepackage{booktabs}%
\usepackage{algorithm}%
\usepackage{algorithmicx}%
\usepackage{algpseudocode}%
\usepackage{listings}%
\usepackage{svg}
\usepackage{subcaption}
\usepackage{caption}
\usepackage{bbm}
\usepackage{float}
\begin{document}
\title[Article Title]{Causal Entropy, Control and Leadership Dynamics}

\author[1]{\fnm{Sam} \sur{Turley}}\email{Sam.Turley.1@warwick.ac.uk}

\author*[2]{\fnm{Matthew} \sur{Turner}}\email{M.S.Turner@warwick.ac.uk}

\affil*[1]{\orgdiv{Warwick Mathematics Institute}, \orgname{University of Warwick}, \orgaddress{\street{Coventry}, \postcode{CV4 7EZ}, \country{UK}}}

\affil[2]{\orgdiv{Department of Physics}, \orgname{University of Warwick}, \orgaddress{\street{Coventry}, \country{UK}. https://orcid.org/0000-0002-3441-678X}}


\abstract{Collective motion in animal groups provide examples of emergent, decentralised coordination. Here, we examine a bottom-up model of collective behavior based on Future State Maximisation (FSM). In this model agents seek to maximise the diversity of their future visual states over a finite time horizon.
We further assume that a subset of agents have a directional bias, e.g. towards different destinations. We observe swarm fragmentation on increasing (i) the strength of these preferences, or (ii) the difference in preferred directions, or (iii) the number of biased agents. Depending on these factors, biased agents can leave the swarm alone, leaving behind all other agents, or they can entrain some fraction of the group to leave with them. We further study the role of a classical nearest-neighbor alignment term on cohesion. Notably, we identify the existence of an finite, optimal coupling strength that suppresses fragmentation and maximises the flock cohesion. Our results demonstrate that FSM can be successfully combined with classical flocking rules, offering a flexible  framework for modeling intelligent collective systems.}

%%================================%%
%% Sample for structured abstract %%
%%================================%%

\keywords{Collective Motion, Leadership, Entropy Production}



\maketitle

\section{Introduction}\label{Introduction}
When studying collective motion a natural question is whether the decision making is individual or collective. Some animal species have leaders \cite{king2008dominance} whilst others use social hierarchies \cite{nagy2010hierarchical} in which decision making is more distributed. Indeed, collective motion can arise even when there is little or no social structure and animals contribute on an equivalent basis. Starlings are thought to be an example of a social group free of leaders \cite{attanasi2015emergence}. However, it is thought that collective decisions, in which many individuals play a role in the decision making, may have advantages in terms of the processing of global information. There are already many studies exploring collective decision making in various model systems  \cite{ward2008quorum, strandburg2015shared, sampaio2024multidimensional, sumpter2008consensus}, together with frameworks for studying influence and leadership \cite{strandburg2018inferring}. Computational models have also been used to explore leadership, 
%to discuss - this word
``democratic'' decision-making and collective conflict resolution \cite{couzin2005effective, couzin2011uninformed, pinkoviezky2018collective}. Here, agents have an equal influence on the group. More strongly opinionated agents cannot influence the decision more those those who are opinionated. These models all involve some form of co-alignment, together with long-range attraction to generate cohesive collectives and short-range repulsion for collision avoidance. 

One animal system that has been extensively studied is the large flocks of starlings, known as ``murmurations''. These birds flocks can contain many thousands of individuals and often occur either as the birds leave their roosts in the morning or return to roost for the night. It has been shown that starling flocks have scale-free velocity correlations \cite{cavagna2010scale}, which are otherwise thought to be rare in nature. Other properties include locally super-diffusive motion \cite{cavagna2013diffusion}, with information transfer that is also super-diffusive. This means that animals use signals that are faster than relying on a diffusive neighbour-to-neighbour consensus \cite{attanasi2014information}. Some evidence of local leadership have been observed, although this leadership is a consequence of their location, rather than social rank. It has been shown that sporadic turns are initiated by birds located on the border of the flock, and that they do not remain leaders after the turn \cite{attanasi2015emergence}. Another quality that seems to emerge in starling flocks is the state of marginal opacity. Here, the flock regulates its density to a state that allows substantial line-of-sight through the crowded environment, while maintaining a highly cohesive group \cite{pearce2014role}. 
%However, its difficult to confirm this using 3D particle tracking, due to sampling biases (observations would frequently be obscured above marginal opacity). 
Many models study equations of motion in which the equivalent of forces or torques can be identified that drive changes in orientation or speed \cite{chen2014minimal, peruani2017hydrodynamic}. However, these models do not naturally target marginal opacity as it requires parameter (re)tuning, depending on the number of agents $N$. Parameter fitting for starling flocks is challenging, since existing 3D data capture techniques employ multiple synchronised cameras \cite{ballerini2008interaction}. While this data has been seminal in providing understanding the phenomenon of bird flocks it is not without limitations. In particular, in order to simultaneously track the positions of agents using video from different perspectives requires the flock to be not more dense than a state of marginal opacity. This is because, at densities higher than this, the video tracking will be frequently obscured \cite{ballerini2008interaction}. Despite the quantitative nature of the FSM model being similar to starling flocks, we are not exclusively focused on this system and we use the terms ``swarm" ``and collective" throughout. 
% I wonder about the emphasis on marginal opacity here. Are we studting/invoking this in the current work. Its fine to mention it but this seems like possible overkill.
%ST I potentially agree. This was background reading and notes from the literature moulded into paper-like text. If we are being general, this may be too much about Starlings and I am okay to chop. Conversley, if we submit with it and it is recommeded to be removed, we do it then. Unless you beleive that this is most likely regardless and the paper should be "lean and mean".
One model that supports the marginal opacity property uses the visual projection of the collective for directional targeting with some nearest-neighbour alignment \cite{pearce2014role}. More recently, ``bottom-up'' models working with causal path entropies to model reorientation also produce marginally opaque collectives with no extrinsically imposed density regulation \cite{charlesworth2019intrinsically, devereux2023environmental}. These works propose the principle of Future State Maximisation (FSM) and focus on intelligent decision-making in complex environments \cite{wissner2013causal, Ebmeier2017path}. In this study we use a variant FSM model that is adapted to include leadership, with some agents ``informed", in our case those that can identify and target a preferred direction. This is  similar to earlier research \cite{couzin2005effective, couzin2011uninformed} exploring questions of leadership and group response with collective particles. However, it is not clear if the results from Viscek-style models carry over to this class of bottom-up model. 

\section{Methods}\label{Methods}

We first re-introduce the FSM algorithm for self-propelled particles for clarity. Our model is a variant of those in the literature \cite{charlesworth2019intrinsically, devereux2023environmental} with the crucial difference that here we incorporate differential decision making in which agents can act according to different rules. This requires us to reformulate the model in terms of a quantity that can be identified as a free energy, instead of one that relies exclusively on a (future) entropy that would be computed in the same way for all agents. We study $N$ agents in 2D space. These agents are circular discs with unit radius, $b=1$, which therefore defines the single characteristic  length scale for the problem. For simplicity, we analyse agents that are ``phantom", meaning they are able to overlap and pass through each other without explicit resistance although the algorithm does assign a form of penalty to collisions, discussed below. 

\begin{figure}
    \centering
    \includegraphics[width=\linewidth]{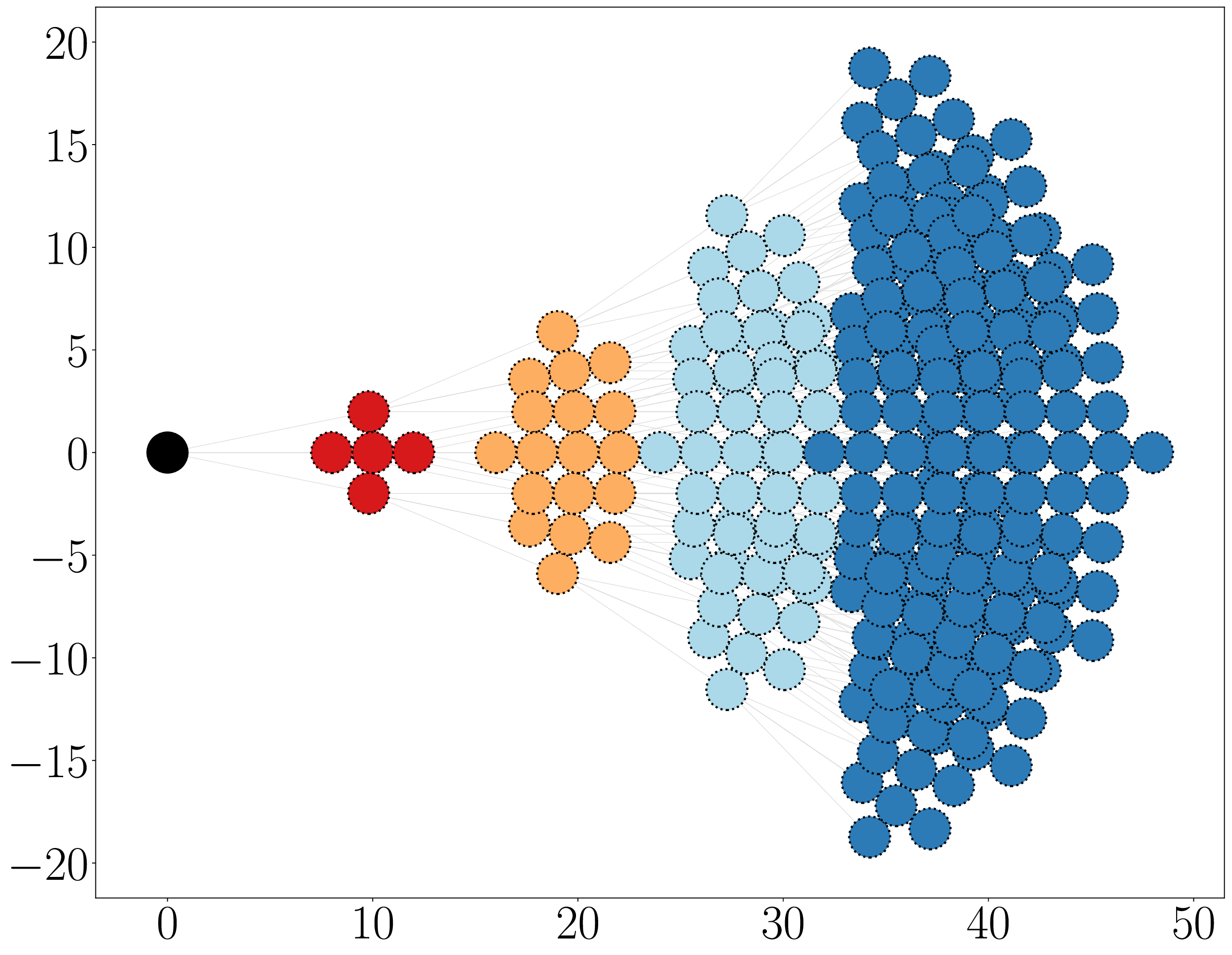}
    \caption{{\bf A visual representation of the movement template.} An agent, shown as a black disc, and its accessible positions in the future, as show in varying colours. The red discs are one action away, with other colours denoting more actions required to reach each position. Note that the orientation at each node is also important, since it affects how the the subsequent trees grow.}
    \label{fig: templete}
\end{figure}

At time $t$, the $i^{\rm th}$ agent arrives with velocity 
\begin{equation}
    \bf{v}_i^t=v_i^t
    \begin{pmatrix}
        \cos{\theta_i^t}\\
        \sin{\theta_i^t}
    \end{pmatrix}
\end{equation}
Thus, $\theta_i^t$ is the orientation angle in the 2D space with respect to some arbitrary axis and $v_i^t$ is the scalar speed at time $t$. 
The updated position at the next timestep is then
\begin{equation}
    \bf{r}_i^{t+1}=\bf{r}_i^{t}+\bf{v}_i^{t+1}.
\end{equation} Thus we use unit time steps throughout. At each time $t$ the agent adopts one of five possible actions, $a \in \mathcal{A} = \{a_l, a_r,a_s,a_0,a_f\}$. These change its speed and/or orientation in the forthcoming timestep $t+1$. The baseline action, $a_0$, has the agent moving with nominal speed $v_0$ and continuing in its current direction, $\theta_i^t$. Alternatively, the agent can move {\underline{s}}lower or {\underline{f}}aster, $a_{s}, a_{f}$, with speed $v_i^{t+1}=v_0 \pm \Delta v$ respectively, while continuing in its current direction. Finally, the agent may reorient either {\underline{l}}eft or {\underline{r}}ight with $a_{l}, a_{r}$ involving $\theta_i^{t+1} =\theta_i^t\pm \Delta \theta$ at speed $v_0$. See Fig. \ref{fig: templete} for a graphical representation of these actions and how sequences of actions generate a tree of positions and orientations accessible in the future.

At this point, the model already has a number of adjustable parameters, $v_0$, $\Delta v$, $\Delta\theta$ as well as $N$, the number of agents. A natural, but by no means unique, choice for the reorientation angle $\Delta\theta$ is one in which the positional changes in the next timestep are nearly isotropic - the agents can move sideways as much as they can move forwards or backwards, relative to the position generated by the nominal action $a_0$. Although this is an arbitrary assumption it does provide a convenient definition that reduces the free parameters by one.
\begin{equation}
    \Delta \theta = 2\arcsin\left(\frac{\Delta v}{2v_0}\right).
\end{equation}

In what follows we also choose $v_0 = 10$, corresponding to relatively fast moving agents:  moving ten times their size between reorientation decisions. To ensure that the agents are always relatively fast we restrict ourselves to modest speed variation $\Delta v=2$. For these choices $\Delta \theta \approx 0.2\approx 11^\circ$. These actions generate a spatial template for the positions accessible at $t+1$. Thus, repeated application of this template generates positions accessible at $t+2$, $t+3$ etc. This creates a branching tree of possible future positions, see Fig. \ref{fig: templete}. The tree shows all the possible positions over some time horizon. In what follows we adopt a tree depth of $\tau=4$ timesteps.

Each agent is equipped with the capability to observe its surroundings. The simplest version of this perception is one in which the agents can differentiate only between lines-of-sight that terminate on another agent, perceived as ``dark'', or those that extend indefinitely (``to the sky''), perceived as ``light''. To describe the environmental state of each agent at each timestep we use the footprint generated by a projection of the collective onto the agent \cite{pearce2014role}. This projection involves a finite number of angular sensors, each recording information on whether the line-of-sight in that direction registers another agent or not.

\begin{figure}
\centering     %%% not \center
\includegraphics[width=\linewidth]{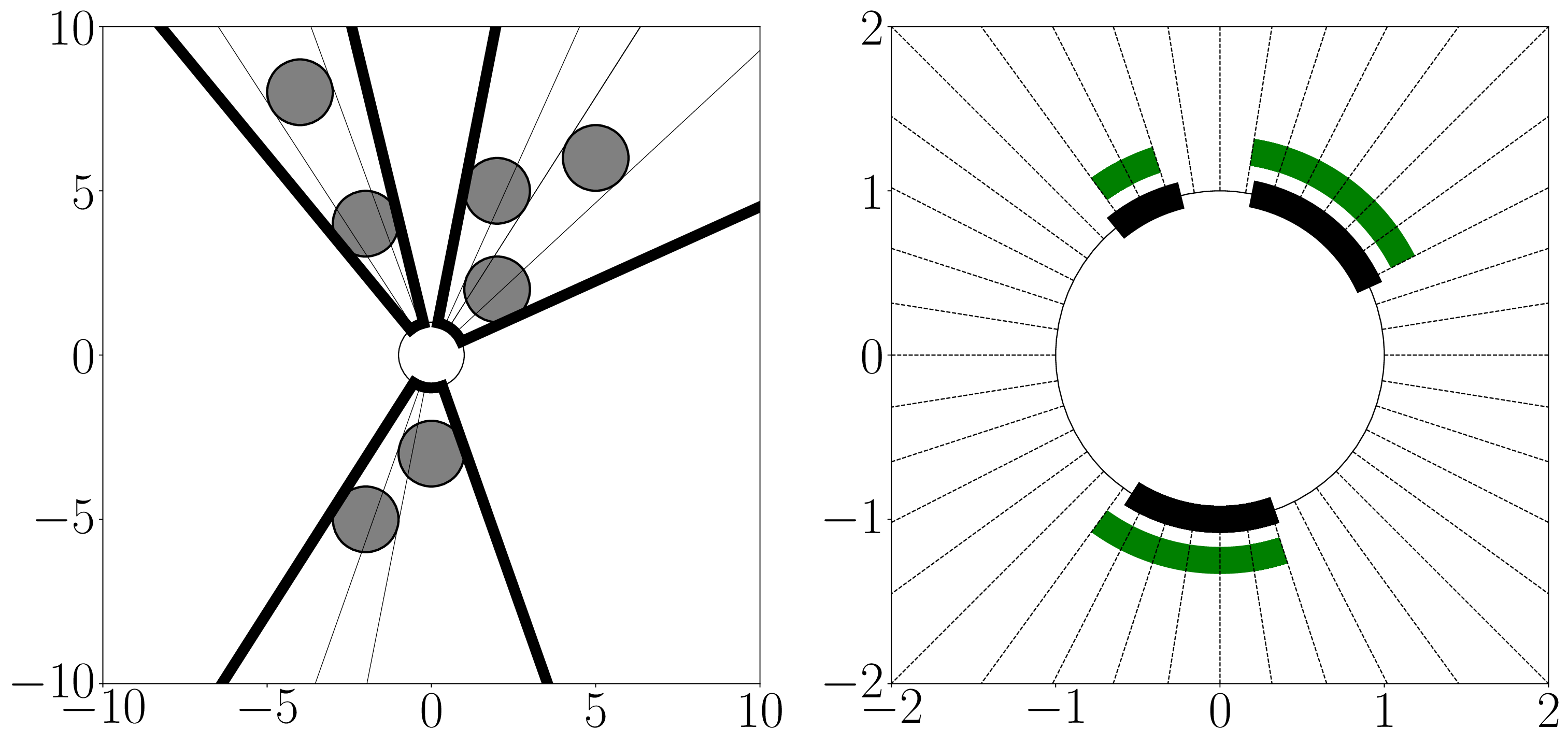}

\caption{{\bf A sketch showing how the visual state vector is defined. } 
On the left is an example swarm configuration, as observed by the central white agent $i$. The thick lines represent the boundaries between open and obscured vision outwards. These boundaries are the union of the projections of the other agents, shown by the thin lines. On the right, we see the $2\pi$ input space surrounding agent agent $i$ is split into $n_s = 40$ equally-sised sensors, denoted by the dashed lines. The information that arrives at the sensors is completely defined by these boundaries of the projection. A sensor is considered active if more than half of the interval is obscured. The thick, black arcs show the visual footprint from the swarm, and the green arcs demonstrate which sensors are consequentially active.}
\label{Fig:2}
\end{figure}

To understand this procedure see Fig. \ref{Fig:2}. Here we show a central agent $i$ onto which we have radially projected the other $N-1$ agents inwards, onto the surface of agent $i$. This involves taking pairs of lines tangential to the edge of each agent $j$ that pass through the centre of $i$ - in the region between any such pair the visual sensor(s) register dark, while regions outside all such pairs of lines register light.

Note that some agents may be partly occluded but still produce regions of dark. The union of these dark intervals provides a visual state for each agent $i$ at time $t$. We discretise the full angular sensor space, extending over an angle $2\pi$, into $n_s$ equally sised {\underline{s}}ensors.
Each sensor is considered dark if more than half of the sensor interval is dark. One can then construct an $n_s$-dimensional vector with elements that are the individual sensor states. We call this the visual state for agent $i$ at time $t$ and
denote it as $\boldsymbol{\psi}_i^t$. Alternatively, we can index specific sensors using $\kappa$, i.e the $\kappa$-th sensor is $\psi_{i,\kappa}^{t}$.
In what follows we take $n_s = 40$ \cite{charlesworth2019intrinsically}.

We now need to define how the agents take actions in the present. In simple terms, agents compare possible actions in the present and choose the action that provides access to the greatest variety of visual states in the near future. 
In order to do this, we need to understand how actions in the present affect visual states in the future. Each agent generates a prediction of its own position and those of all other $N-1$ agents into the future, up to a time horizon at $t+\tau$. The $i^{\rm th}$ agent has perfect control over its own actions, and these actions deterministically generate the future positions, see Fig. \ref{fig: templete}. 

However, each agent does not have control over the other agents, nor the positions that they will occupy in the future. In order to estimate these positions, each agent employs a heuristic. The simplest heuristic is to assume that other agents will move with nominal speed $v_0$ and maintain their current orientation $\theta_j^t$. We refer to this as the ballistic heuristic. While it may sound crude, it is a good approximation, given the high degree of co-alignment generated in the model of interest to us.

\begin{figure}
\centering     %%% not \center
\includegraphics[width=\linewidth]{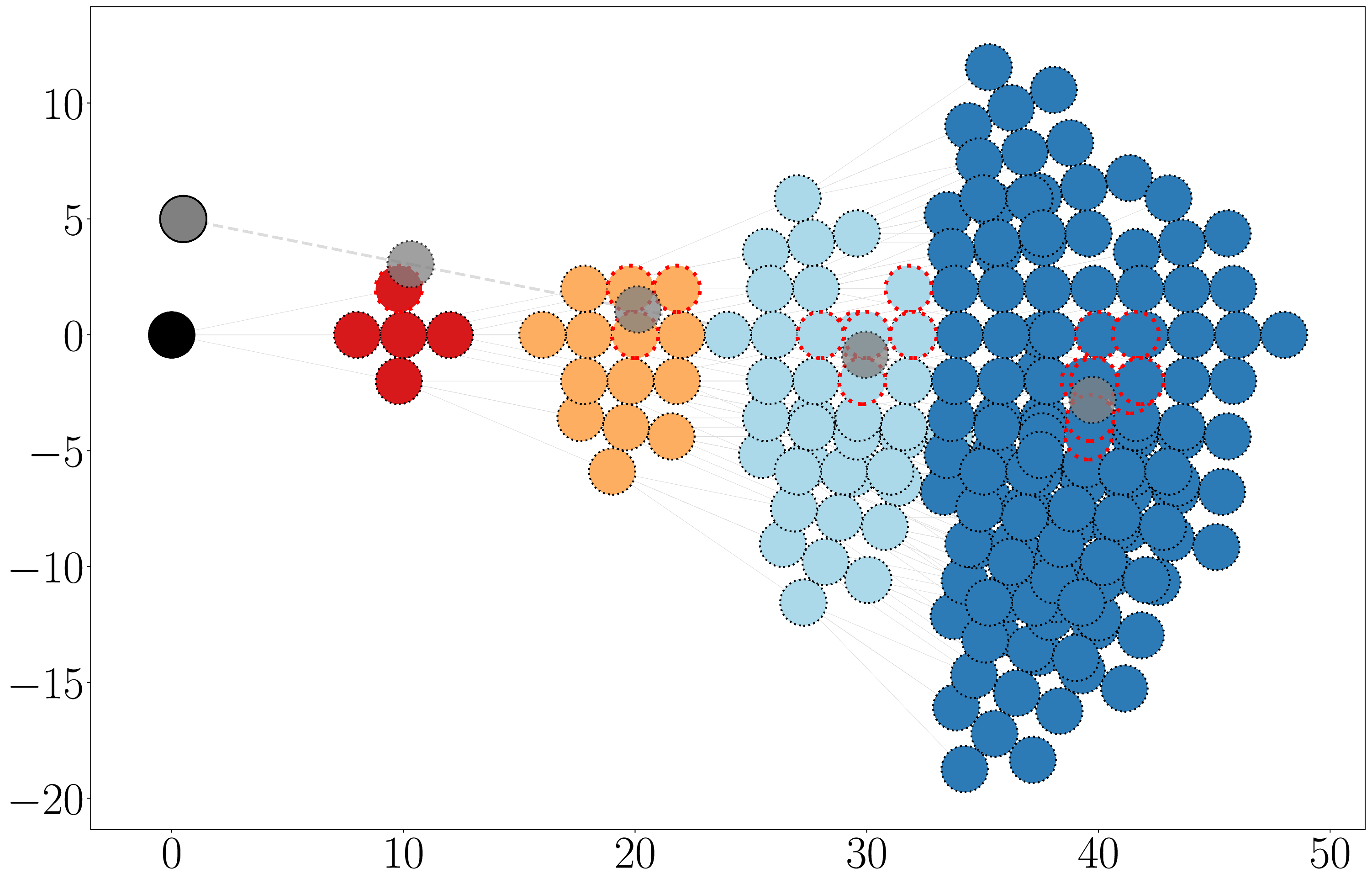}
\caption{{\bf A visual representation to show the affect of truncation to due predicted collisions.} At each safe node, a visual state is taken. A node is safe if there is no collisions with other agents prior to the decision. But how does an agent know if there will be a collision? To simplify computations, agent $i$ assumes all other agents are ballistic, travelling forwards linearly at nominal speed, maintaining their orientation. Here the grey disc is projected to collide with some of the decision nodes of the tree, indicated by a dotted red outline. A collision means that a visual state is not recorded and further actions are not explored. This means the tree grows asymmetrically, favouring ``safer" root actions. }\label{Fig: plotBallistic1}
\end{figure}

Furthermore, to suppress collisions, we prune the action tree: if an action would lead to a collision (at any continuous time between re-orientation actions), the tree is terminated, see Fig. \ref{Fig: plotBallistic1}. The tree grows in directions that favour non-collision actions. These non-truncated branches have more opportunity to observe variations in visual states with a correlation between (lack of) truncation and likelihood of being selected as the preferred action. 

We need to formally define what is meant by ``variety of visual states". We shall use a tilde to distinguish between hypothetical actions and visual states used for calculation and the actual actions and visual states of the model. Firstly, we denote a sequence of hypothetical actions of length $\tau$ by 
$\tilde{\mathbf{a}}^t_i =\{\tilde{a}_i^t,..., \tilde{a}_i^{t+\tau-1}\}$. We index a specific hypothetical action in this sequence by $\tilde{\mathbf{a}}^{t, d}_i$, where $d$ can be thought of as depth or delay. Occasionally, a decision node would cause a collision regardless of action selected. If so, we consider this to be a terminal node. We ensure the path remains of length $\tau$ by using a placeholder action, $a_\text{term}$, to complete the path. 

After each hypothetical action is taken, a hypothetical visual state is recorded, where it is assumed that all other agents move with the ballistic heuristic. This visual state is denoted $\widetilde{\boldsymbol{\psi}}$, which can also be indexed using $d$. Thus each path generates a sequence of visual states $\boldsymbol{\Psi}^t_i(\tilde{\mathbf{a}}^t_i) = \{ \widetilde{\boldsymbol{\psi}}_i^{t+1},..., \}$. 
We bundle the possible paths into sets, or subtrees, which are associated with the root action $a_i^{t,0}$, denoted $\mathcal{B}^t_i(a)$. A path belongs to a subtree, or ``branch" if and only if $\tilde{a}_i^{t,0} = a$. Due to the branching nature of the paths, we have $\vert \mathcal{B}^t_i(a) \vert \leq \vert \mathcal{A}\vert^{\tau -1}$. We can now measure the variation of visual states associated with a root action $a$. Along every path within a bundle, we record the number of occurrences of each visual state, denoted $\mathcal{N}_i(\boldsymbol{\psi}', a)$.
\begin{equation}
\mathcal{N}_i(\boldsymbol{\psi}', a) = 
   \sum_{\tilde{\mathbf{a}}^t_i \in \mathcal{B}^t_i(a)}
    \sum_{d=0}^{\tau - 1}
    \mathbbm{1}(\tilde{a}^{t,d}_i \ne a_\text{term})
    \mathbbm{1}(\boldsymbol{\psi}_i^{t,d} = \boldsymbol{\psi}'),
\end{equation}
where $\mathbbm{1}$ is an indicator defined by
\begin{equation}
    \mathbbm{1}(\boldsymbol{\psi_i^{t,d}} = \boldsymbol{\psi}')
    :=
    \prod_{\kappa=1}^{n_s}  \mathbbm{1}(\widetilde{\psi}_{i, \kappa}^{t, d} = \psi_\kappa').
\end{equation}
By using paths, rather than nodes, we count non-leaf nodes multiple times. This bias towards internal nodes is natural, as future states are less certain and therefore are more discounted. From the raw count of each visual state, we can normalize to calculate the proportion of occurrences,
\begin{equation}
\mathbb{P}_i(\boldsymbol{\psi}, a) = \frac{\mathcal{N}_i(\boldsymbol{\psi}, a)}{{\sum_{\boldsymbol{\psi}'}}\mathcal{N}_i(\boldsymbol{\psi}', a)}.   
\end{equation}
From this distribution, we calculate the Shannon Entropy \cite{shannon1948mathematical} of an action:
\begin{equation}
\mathcal{S}_i^t(a) =
-\sum_{\boldsymbol{\psi}} \mathbb{P}_i( \boldsymbol{\psi}, a) 
\ln \big[ \mathbb{P}_i(\boldsymbol{\psi}, a) \big].
\end{equation}
The action selection policy of FSM is then given by 
\begin{equation}
a_i^t = \text{argmax}_{a \in \mathcal{A}} \mathcal{S}_i^t(a).
\end{equation}
It has been shown that the FSM models yields collective motion that is cohesive, highly co-aligned, and displays other real-world characteristics, including scale-free correlations and marginal opacity \cite{charlesworth2019intrinsically,devereux2023environmental}.

We now explore implement leadership into the FSM framework. Therefore, we need some method for the agents to express a directional preference. This is achieved by altering the action selection policy. We pose this as an energy minimisation problem where each agents minimizes its instantaneous $E_i$,
\begin{equation}
    E_i = \mathcal{U}_i - T \mathcal{S}_i,
\end{equation}
where $\mathcal{U}_i$ is the internal energy associated with directional preference, $T$ is the temperature, here set to unity $T=1$, and $\mathcal{S}_i$ as the path-entropy from FSM. For the internal energy $\mathcal{U}_i$ we use a magnetisation-like energy,
\begin{equation}
\mathcal{U}_i = -\omega_i (\textbf{v}_i^{t+1}(a) \cdot \textbf{M}_{i}),
\end{equation}
where ${\omega}_i$ is a coupling strength, $\textbf{M}_{i}$, is a unit magnetisation vector and $\textbf{v}_i^{t+1}(a)$ is the ``spin'' direction of the agent $i$ after the action $a$. In simple terms, the internal energy for agent $i$ is minimised when aligned with its directional bias vector $\textbf{M}_{i}$. 

The vector $\textbf{M}_i$ can be chosen in many different ways; it can be a fixed direction, like a global magnetic field, or it can be the direction vector towards a target location etc. Alternatively, it could be the direction vector produced by any classical co-alignment mechanism. For example, this could be the consensus orientation direction implied by a local Viscek model or a global Cucker-Smale model \cite{vicsek1995novel, cucker2007emergent}. 

\begin{figure}
    \centering
    \includegraphics[width= \linewidth]{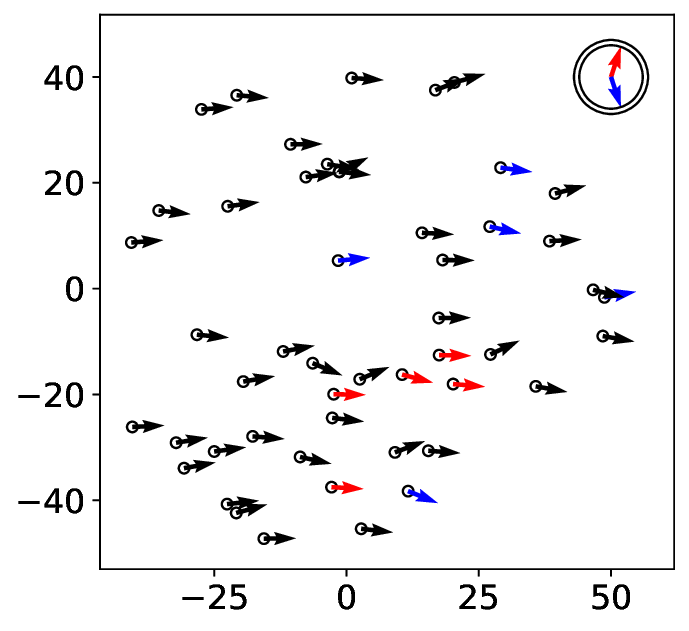}
    \caption{{\bf Initial configuration with informed (biased) agents in colour.} $N=50$ agents are initialised in infinite 2D space. Here 10 agents, in two groups of $n=5$, are informed (biased) and have a directional preference as shown in the compass in the upper right: red/blue agents have an alignment bias to the red/blue direction respectively.}
    \label{fig:SetUp}
\end{figure}

The new action selection policy that we study in this paper is then given by
\begin{equation}
    {a_i}^t = \text{argmax}_{a \in \mathcal{A}} \left [ {\omega}_i  ({\textbf{v}}^{t+1}_i(a) \cdot \textbf{M}_i) + \mathcal{S}_i^t(a) \right ].
\end{equation}
%It is not required for that agents have the same $\omega_i$ or $\textbf{M}_i$. 
Agents that are not affected by the magnetic field, $\omega_i = 0$, are uninformed agents, since they do not have a directional preference. We initialize a collective of $N = 50$ agents randomly located within a disc of radius $R = N$, with orientation $\theta_i = \eta_i \Delta \theta $, where $\eta_i \sim \mathcal{N}(0,1).$ We take a group of $2n$ informed (biased) agents, with indices $i \in \mathcal{I}$, leaving a total of $N - 2n$ uninformed agents. There is no physical distinction between informed and uninformed agents: agents cannot tell which other agents have a directional preference. To simplify the model we take $\omega_i = \omega$: all informed agents have the same coupling strength. To preserve symmetry we split the agents into two groups of $n$ agents. We set the preferred directions of each group to be mirror images at an angle of $\pm \Theta$ to the x-axis, with one sub-group aligning with a vector $\textbf{M}_\Theta$ and the other with $\textbf{M}_{-\Theta}$. Fig. \ref{fig:SetUp} shows an example of an inital configuration. 

\section{Results}\label{Results}

\subsection{Magnetisation Energy}

We study $N = 50$ agents, with five different $\omega$ values and five different field orientations $\Theta$. Averaging over 50 simulations, which run up to $t_\text{max} = 500$ time steps. We examine the extent to which the collective stays intact, to what extend it splits into smaller groups and by extension, the ability of informed agents to lead the rest of the collective. Since we are interested in swarm fragmentation, we need a measure of the clustering of the agents. We use the DBSCAN algorithm \cite{Ester_1996}, which allows us to measure the number of distinct clusters and hence track fragmentation of the collective. Broadly, agents are in the same cluster if they are spatially close and are co-aligned. More details on our specific implementation of DBSCAN are given in Appendix \ref{secA1}. We denote the group containing agent $i$ by $C_i$. We calculate the average cluster size of $C_i$ where agent $i$ is informed, $i \in \mathcal{I}$:
\begin{equation}
    \bar{C} = \frac{1}{\vert \mathcal{I} \vert} \sum_{i \in \mathcal{I}} \frac{\vert C_i \vert}{N}
\end{equation}
This gives a simple and robust measure of the splitting that occurs. A value of 1 implies that the whole collective is one cluster, since the informed agents are in a cluster with all $N$ agents. A value of 0.5 suggests that the informed agents are, on average, in collectives with half of all the agents. We expect this value to occur when the two groups of informed agents lead the collective in their respective preferred decisions. Smaller values suggest that the clusters containing the informed agents are a small proportion of the collective, i.e the informed agents are moving nearly alone or with each other.

\begin{figure*}
    \centering
    \includegraphics[width = 0.95\linewidth]{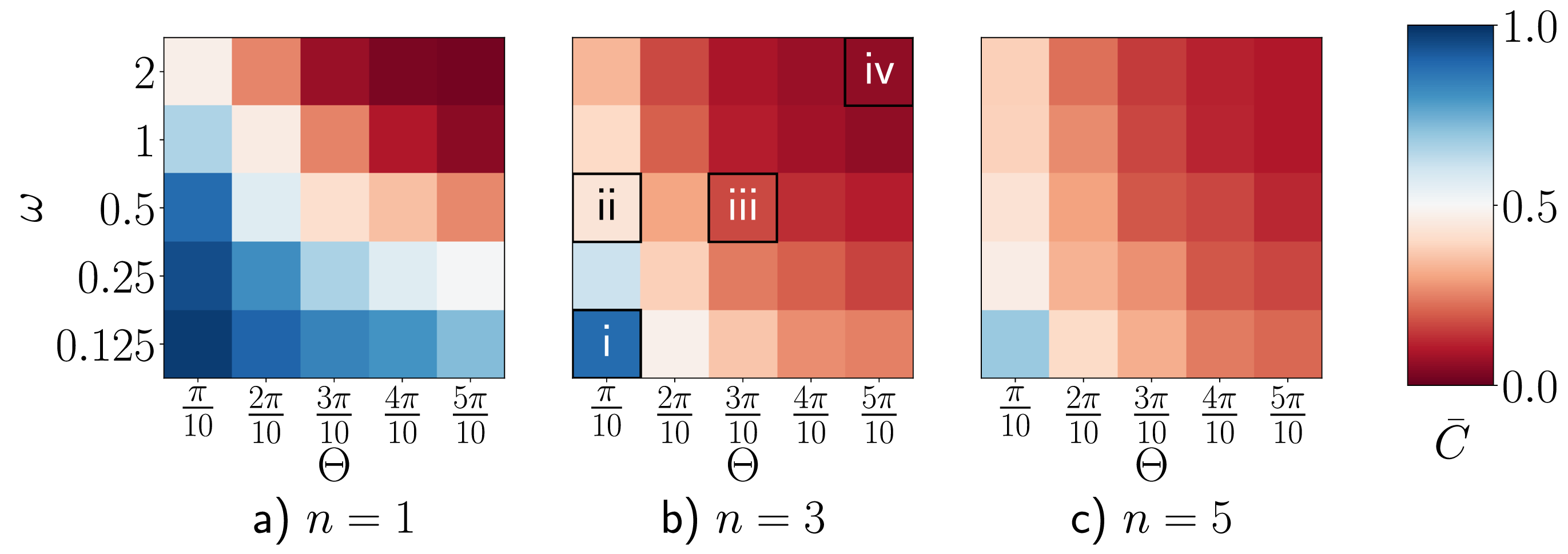}
    \caption{{\bf Heat maps showing the average cluster size of the informed agents, $\bar{C}$.}  The axes show the preferred angle, $\Theta$, against the directional strength, $\omega$. The splitting proportion, $\bar{C}$, is monotonic in both directions. For $n=1$, small $\omega$ and $\theta$, the whole collective stays together: the informed agents do not have enough influence over the group to cause splitting. The top right of each graph demonstrates the extreme effect of the directional preference; the FSM entropy is not enough to keep the informed agents within the collective and they leave regardless of whether other agents follow. Along the negative diagonals, we see a similar shading the same colour, suggesting that larger differences in directional preference require weaker strength to ensure the collective stays together. Squares with roman numerals are explored further below in Fig. \ref{Fig: Ridgelines}. The values at each location are averaged over time steps 490-500 for collectives of $N = 50$ agents.}  
    \label{fig:heatmaps}
\end{figure*}

\begin{figure*}
    \centering
    \includegraphics[width=0.95\linewidth]{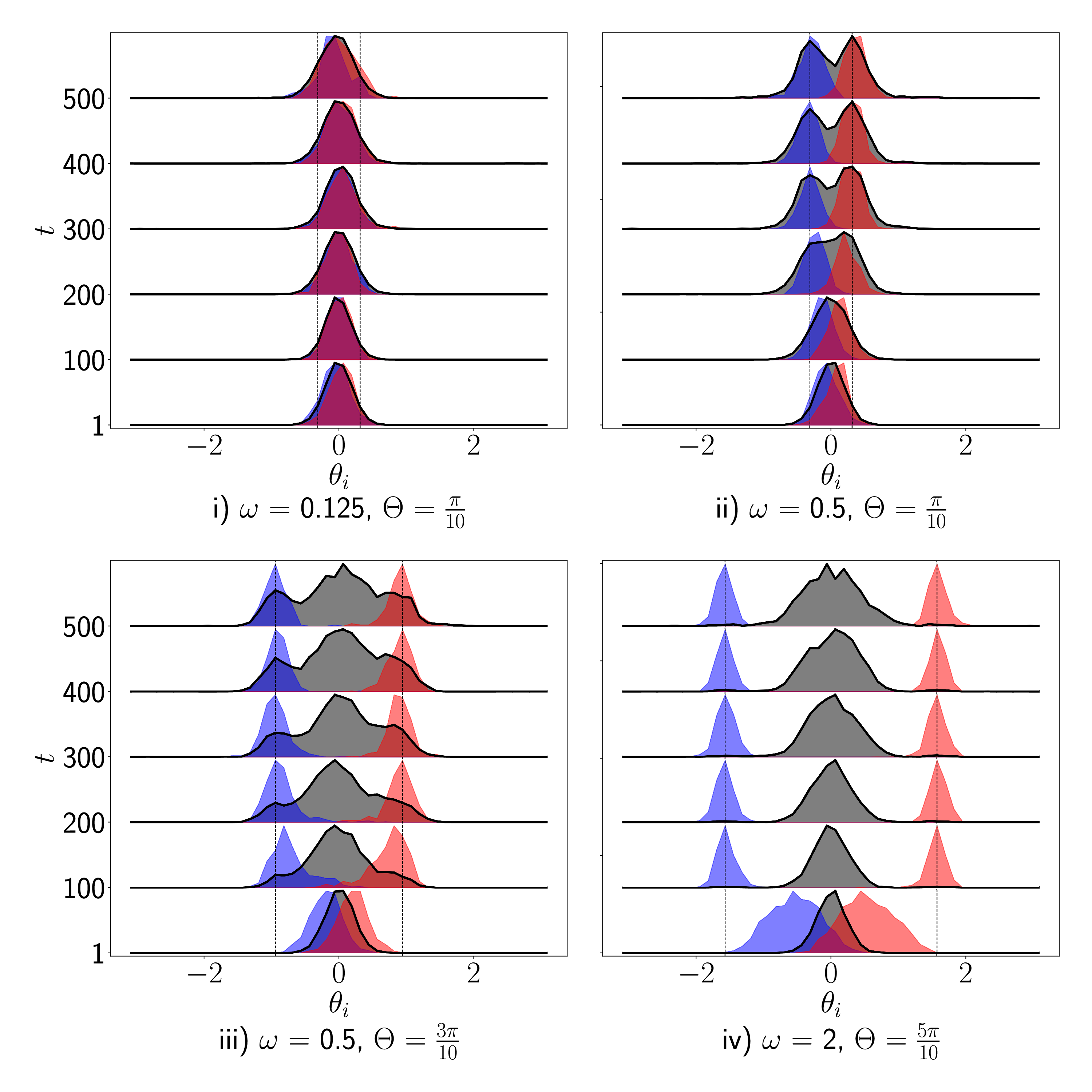}
    \caption{Four probability density histograms of all agents' orientations over time for different set ups, where orientations are averaged over 10 time steps. The vertical dashed lines represent the target orientations. In (i), we can see that the informed agents cannot split the flock into two. In (ii), we see a symmetric split, where we have two distinct clusters of agents centered on the target orientations. In (iii), we see a tri-modal distribution of uniformed agents. Two peaks are associated with the informed agents and a singular peak at $\theta_i =0$ shows that some agents have not been convinced to change orientation. This may be due to the fragmentation process being fast. In (iv), we see the extreme case where the informed agents leave the collective, regardless of the uninformed agents.}
    \label{Fig: Ridgelines}
\end{figure*}

The results for different numbers of informed agents $n = 1, 3$ and 5 with increasing $\omega$ and $\textbf{M}_\Theta$ are shown in Fig. \ref{fig:heatmaps}. The broad spectrum of $\bar{C}$ demonstrates the existence of various types of splitting events. If the preference strength $\omega$ is small or the difference between the current heading and the directional preference small, then the collective does not split. This is seen by the large values of $\bar{C}$ in the bottom-left of the each heatmap. The informed agents' entropies are maximised by staying within the collective, since the internal energy to change, $\mathcal{U}$ is small relative to $\mathcal{S}$. As $\mathcal{U}$ increases, by increasing either $\omega$ or $\theta$, fragmentation increases and consequently we see a decrease in $\bar{C}$. We see a clear trend along the negative diagonals, where the value of $\bar{C}$ is constant. This means that a strong preference for a small change has the same effect on fragmentation as a weak preference for a large change. The informed agents maximize their entropy by balancing remaining with the collective and heading in their preferred direction. The rate at which informed agents attempt to head in their preferred direction affects whether the uninformed agents follow or stay. Uninformed agents' decisions are only based upon the FSM entropy. This is generally maximised by staying within the collective as there is better access to distinct visual states. For intermediary values of $\mathcal{U}$, the informed agents leave the flock after ``persuading" some uninformed agents to travel in the same direction. In the top right, we observe the smallest values of $\bar{C}$. This is due to the internal energy playing a dominating role in the action selection. The informed agents' entropies are maximised by leaving the collective, regardless of the uniformed agents. These agents ``need" to go in their preferred direction. In general, we see that increasing the number of informed agents increases the fragmentation of the collective. For $n=5$ informed agents, a significant proportion of the flock has a directional preference, so the realisation of achieving a change of heading is easier. We can observe the behaviour of the flock by looking at the agents' orientations $\theta_i$ over time. The different phenotypical outcomes of fragmentation are shown in Fig. \ref{Fig: Ridgelines}.

\subsection{Nearest Neighbour Alignment}

We can also extend the internal energy further. To mitigate the fragmentation of the collective, we may amend $\mathcal{U}$ to include any classical or modern flocking methods. To demonstrate this, we include a nearest-neighbour alignment for all agents, informed and uninformed. It is claimed that topological alignment (as opposed to distance-based) is highly important for starling flocking and that between six or seven is the critical number of neighbours \cite{ballerini2008interaction}. Scaling this for a 2D system, we use four nearest-neighbours. We denote the neighbours for agent $i$ by $\langle ij \rangle$ and our alignment is that of a spin model, inspired by the XY model with coupling strength $J$ \cite{stanley1968dependence}. The action policy becomes:
\begin{align}
    a_i^t =&  \text{argmax}_{a \in \mathcal{A}}  \\
    & \left[ \underbrace{J \sum_{\langle ij \rangle} \textbf{v}_i^{t+1}(a) \cdot \textbf{v}_j^{t} + \omega  ({\textbf{v}}_i^{t+1}(a) \cdot \textbf{M}_i)}_{\mathcal{U}} + \mathcal{S}_i^t(a) \right]. \notag
\end{align}

\begin{figure*}
    \centering
    \includegraphics[width = 0.95\linewidth]{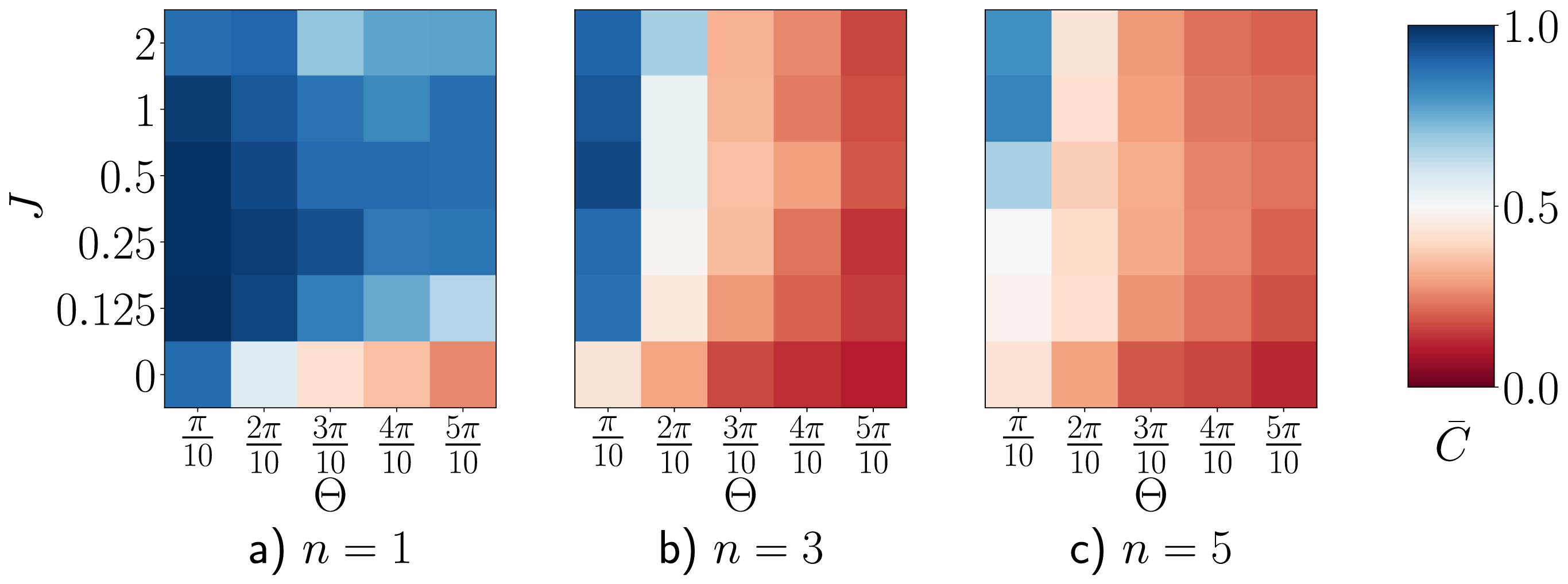}
    \caption{{\bf Heat maps showing $\bar{C}$ with additional spin coupling.} Heat map values are averaged over time steps 490-500 with fixed splitting strength $\omega$ = 0.5 and varying spin coupling $J$ for $n = 1,3,5$. The y-axes are the strength of the nearest-neighbour coupling, $J$. Reading horizontally the heat maps shows the same pattern as before, increasing the angle increases the amount of fragmentation. One might expect that increasing the nearest-neighbour coupling would monotonically increase the swarm's cohesiveness. However, reading vertically across the heat maps suggests this is not the case. There appears to be an optimal nearest-neighbour coupling to maintain a singular collective with $\omega = 0.5$.}
    \label{fig:Spinning Heatmaps}
\end{figure*}

Results with fixed $\omega = 0.5$ and varying $J$ and ${\bf M}_\Theta$ are shown in Fig. 
\ref{fig:Spinning Heatmaps}. The bottom row on the heat maps, given by $J=0$, is the middle row taken from Fig. \ref{fig:heatmaps}, as a reference. We see that including the spin coupling greatly increases the cohesion across the across the different scenarios. This is due it being more entropically favourable for agents to travel in the average direction of neighbours, as opposed to only attempting to reposition with a collective. One can imagine that when an agent or group attempts to leave the collective, the uninformed agents now form a connection between the original swarm and the those attempting to depart. Rather than quickly deciding which swarm to join, the result is a splay of agents, providing more time for the collective to be directed or for the group to return. As such, the collective is more robust against noisy decisions or losing numbers due to extreme preferences.   

A key observation from the heat maps is that there appears to be a non-monotonicity with respect to $J$. Increasing $J$ from can increase cohesitivy, but increasing it too much, such that this term dominates the action selection policy, can reduce cohesivity slightly. This is because nearest-neighbour alignment does not explicitly promote cohesion; in the FSM model, clustering is provided by the FSM entropy term. Thus, if $\mathcal{S}$ is only a small contribution when calculating $\text{argmin } E$, the collective may fragment into multiple co-aligned flocks. 

\section{Discussion}\label{Discussion}
When thinking about real-world systems, such as starling murmurations, we would expect individual members to balance the competing goals of maintaining flock cohesion and travelling in their prefered direction. One could imagine a bird subtly altering its flight pattern to demonstrate a preferred direction, even if only temporarily. How such behaviour is tuned to maximize preference expression is not immediately obvious. However, our results indicate that a more significant direction change from the collective's heading requires weaker expression to maintain collective cohesivity. Agents with strong directional preferences leave the collective rapidly and forgo the benefits of being within the collective. Furthermore, this tendency is exacerbated when multiple agents share the same preference; these small groups fragment from the collective together. We can infer that for a given directional preference and number of informed agents, there is a critical $\omega$, under which the swarm remains cohesive and over which fragmentation will occur. We can imagine an optimal ``pushiness" that convinces the other agents to change orientation without fragmentation.
% ST I am using in the sense of an environment where only ones opinions are heard, which re-enforces your own opinion. I am happy to change it if there is something better/ more scientifically appropriate but I do feel the phrase describes the sitution well. 
% Head in prefered direction, so do others with the same preference -> aligning with them -> More strongly re-enforces the prefered direciton -> FSM (we predict they will go straight) PLUS alignment PLUS preference
% Maybe something like "Compounding" or similar works? The fact there are multiple affects to ensure that direction 
In our model, we observe that agents with the same preferences collect together, Fig. \ref{fig:Spinning Heatmaps}. Informed agents shift locally within the collective in their preferred direction; this intra-flock rearrangement is more common in the FSM framework than with force-based model. As such, they are more likely to locate themselves on the border of the collective. This means that informed agents are more likely to cluster with other agents that have the same directional preference. As a result, breaking away from the collective together is more entropically favourable than alone, since a large number of visual states can still be maintained if other agents are leaving the flock too. Nearest-neighbour alignment compounds this phenomenon, as infomred agents' nearest neighbours are often other agents that are share the directional preference.

This work raises an important question about topological alignment. Because nearest-neighbour alignment is scale-free, an individual agent rarely leaves the collective on its own. We see this in Fig. \ref{fig:Spinning Heatmaps}, $n=1$. In contrast, groups of informed agents larger than the topological number of aligning neighbours, are much more likely to break away. This is clear when comparing the heat maps $n=3$ and $n=5$ in Fig. \ref{fig:Spinning Heatmaps}. We are using a 4-neighbour alignment, so it is possible that with $n=5$, all informed agents are each others' nearest neighbours. Instead, with $n=3$, there will always be at least two neighbours that are uninformed. As such, groups smaller than than the topological number of neighbours have a persistent interaction with the collective which can be non-reciprocal when at distance. 

The results seen in the heat maps Fig. \ref{fig:heatmaps}, \ref{fig:Spinning Heatmaps} suggest that cohesion in real-world collectives could be challenging to manage when individuals have different directional preferences. Consider a flock of starlings, deciding on a roost site for the night. Each bird benefits from staying in the group to avoid predation and perhaps for warmth when roosting. Each individual also likely has some preference about roost sites; which trees would be appropriate and which would not. Thus, the collective needs some mechanism for the birds to ``vote", and for individuals to perceive the preferences of other birds. We can imagine that flocking birds may have evolved to learn the group's democratic preference. A bird would need to assess whether small perturbations in others' directions are due to noisy flight patterns or whether they are trying to express a certain preference for roost site. The dance of a murmuration might be the continual voting for where to roost, only halting after enough birds have committed to a direction and the group has found consensus. We see that nearest-neighbour alignment significantly helps cohesion, and this work demonstrates that fragmentation events should be should be rare if expression is mild or preference is insignificant.

Natural extensions of this work might seek to explore classical flocking models within the FSM framework. For example, the magnetisation direction could be changed to a target location, or the sets of informed agents need not be equally sized. Moreover, we could explore the results of changing the terms used in our free energy. For example, we could include neighbour alignment within a fixed radius, or include a long-range attractive force in the model. 

\section{Conclusion}\label{Conclusions}

We analyzed the fragmentation of collectives of intelligent agents, which take actions according to an entropy-production model known as Future State Maximisation (FSM). In this framework, agents sought to keep their future accessible options open over a short time horizon. A small number of agents were also given a directional preference, and thus called ``informed" agents. This was implemented in the FSM framework via an additive free energy term. We found that collective fragmentation increased when the informed agents had a stronger directional preference, $\omega$, or when the angular difference between their preference and the flock's current direction increased. Under extreme conditions, informed agents were even willing to leave the collective alone. Increasing the number of informed agents also increased the likelihood and severity of fragmentation.

To promote cohesion, we introduced a classical nearest-neighbour alignment, in which agents aligned with the four nearest neighbors with coupling strength $J$. We identified the existence of a critical value of $J$ that minimised fragmentation. Below this threshold, alignment was ineffective; above it, strong coupling lead to the FSM entropy contribution being dominated within the action policy. This allowed for group-level departures, as cohesivity was an emergent property of the FSM entropy. 

Most significantly, our findings show that classical alignment models can be integrated with the bottom-up, decision-theoretic FSM framework. This hybrid approach provides a foundation for more sophisticated models that aim to capture the complex properties that intelligent real-world collective system display.

\backmatter

\bmhead{Supplementary information}
\section*{Declarations}
S.T. developed the simulation code, performed numerical experiments, and analysed data.
M.S.T. supervised the project and contributed to development of the model and interpretation of results. Both authors contributed to the writing and editing of the manuscript. Funding was provided by UK EPSRC though the Mathematics for
Real World Systems Centre for Doctoral Training Grant EP/S022244/1. All computational work was carried out using the Scientific Computing Research Technology Platform of the University of Warwick. S.T and
M.S.T. acknowledge the peerless hospitality of Prof. Ryoichi Yamamoto at Kyoto University. All simulation data are available upon request. The authors declare no competing interests.

\begin{appendices}

\section{DBSCAN}\label{secA1}

We use the DBSCAN algorithm \cite{Ester_1996} for calculating clusters, which has two hyper-parameters. The first, $\textbf{Eps}$, is a distance metric, below which two points are labelled in the same cluster. Two points can be in the same cluster despite being further away than $\textbf{Eps}$ if there are intermediary points that connect them. We set $\textbf{Eps} =1$, a scale-free distance based upon the spatial and orientation difference
\begin{align}
d_{i,j}^t &= \frac{1}{2} \frac{\vert \vert  \textbf{x}_i-\textbf{x}_j \vert \vert_2}{\pi \sqrt{N}} \\
&+ \frac{\text{Min}(\vert \theta_i^t - \theta_j^t \vert, 2 \pi - \vert \theta_i^t - \theta_j^t \vert)}{4 \Delta \theta}  \\
\end{align}
The second hyper-parameter is $\textbf{MinPts}$, the minimum number of points required to produce a clustering label. We set this to $\textbf{MinPts} = 1$, as we wish to track whether single agents leave the collective or not.

%% please use the heading ``Extended Data''. %%
%%=============================================%%



\end{appendices}

%% file, and delete the associated \verb+\bibliography+ commands.            %%
%%===========================================================================================%%

\bibliography{sn-bibliography}% common bib file

@article{pearce2014role,
  title={Role of projection in the control of bird flocks},
  author={Pearce, Daniel JG and Miller, Adam M and Rowlands, George and Turner, Matthew S},
  journal={Proceedings of the National Academy of Sciences},
  volume={111},
  number={29},
  pages={10422--10426},
  year={2014},
  publisher={National Acad Sciences}
}

@inproceedings{Ester_1996,
  title={A density-based algorithm for discovering clusters in large spatial databases with noise},
  author={Ester, Martin and Kriegel, Hans-Peter and Sander, J{\"o}rg and Xu, Xiaowei and others},
  booktitle={kdd},
  volume={96},
  pages={226--231},
  year={1996}
}

@article{shannon1948mathematical,
  title={A mathematical theory of communication},
  author={Shannon, Claude Elwood},
  journal={The Bell system technical journal},
  volume={27},
  number={3},
  pages={379--423},
  year={1948},
  publisher={Nokia Bell Labs}
}

@article{charlesworth2019intrinsically,
  title={Intrinsically motivated collective motion},
  author={Charlesworth, Henry J and Turner, Matthew S},
  journal={Proceedings of the National Academy of Sciences},
  volume={116},
  number={31},
  pages={15362--15367},
  year={2019},
  publisher={National Acad Sciences}
}

@article{devereux2023environmental,
  title={Environmental path-entropy and collective motion},
  author={Devereux, Harvey L and Turner, Matthew S},
  journal={Physical Review Letters},
  volume={130},
  number={16},
  pages={168201},
  year={2023},
  publisher={APS}
}

@article{vicsek1995novel,
  title={Novel type of phase transition in a system of self-driven particles},
  author={Vicsek, Tam{\'a}s and Czir{\'o}k, Andr{\'a}s and Ben-Jacob, Eshel and Cohen, Inon and Shochet, Ofer},
  journal={Physical review letters},
  volume={75},
  number={6},
  pages={1226},
  year={1995},
  publisher={APS}
}

@article{chen2014minimal,
  title={A minimal model of predator--swarm interactions},
  author={Chen, Yuxin and Kolokolnikov, Theodore},
  journal={Journal of The Royal Society Interface},
  volume={11},
  number={94},
  pages={20131208},
  year={2014},
  publisher={The Royal Society}
}

@article{ballerini2008interaction,
  title={Interaction ruling animal collective behavior depends on topological rather than metric distance: Evidence from a field study},
  author={Ballerini, Michele and Cabibbo, Nicola and Candelier, Raphael and Cavagna, Andrea and Cisbani, Evaristo and Giardina, Irene and Lecomte, Vivien and Orlandi, Alberto and Parisi, Giorgio and Procaccini, Andrea and others},
  journal={Proceedings of the national academy of sciences},
  volume={105},
  number={4},
  pages={1232--1237},
  year={2008},
  publisher={National Acad Sciences}
}

@article{attanasi2014information,
  title={Information transfer and behavioural inertia in starling flocks},
  author={Attanasi, Alessandro and Cavagna, Andrea and Del Castello, Lorenzo and Giardina, Irene and Grigera, Tomas S and Jeli{\'c}, Asja and Melillo, Stefania and Parisi, Leonardo and Pohl, Oliver and Shen, Edward and others},
  journal={Nature physics},
  volume={10},
  number={9},
  pages={691--696},
  year={2014},
  publisher={Nature Publishing Group UK London}
}

@article{couzin2005effective,
  title={Effective leadership and decision-making in animal groups on the move},
  author={Couzin, Iain D and Krause, Jens and Franks, Nigel R and Levin, Simon A},
  journal={Nature},
  volume={433},
  number={7025},
  pages={513--516},
  year={2005},
  publisher={Nature Publishing Group UK London}
}

@article{strandburg2015shared,
  title={Shared decision-making drives collective movement in wild baboons},
  author={Strandburg-Peshkin, Ariana and Farine, Damien R and Couzin, Iain D and Crofoot, Margaret C},
  journal={Science},
  volume={348},
  number={6241},
  pages={1358--1361},
  year={2015},
  publisher={American Association for the Advancement of Science}
}

@article{ward2008quorum,
  title={Quorum decision-making facilitates information transfer in fish shoals},
  author={Ward, Ashley JW and Sumpter, David JT and Couzin, Iain D and Hart, Paul JB and Krause, Jens},
  journal={Proceedings of the National Academy of Sciences},
  volume={105},
  number={19},
  pages={6948--6953},
  year={2008},
  publisher={National Acad Sciences}
}

@article{couzin2011uninformed,
  title={Uninformed individuals promote democratic consensus in animal groups},
  author={Couzin, Iain D and Ioannou, Christos C and Demirel, G{\"u}ven and Gross, Thilo and Torney, Colin J and Hartnett, Andrew and Conradt, Larissa and Levin, Simon A and Leonard, Naomi E},
  journal={science},
  volume={334},
  number={6062},
  pages={1578--1580},
  year={2011},
  publisher={American Association for the Advancement of Science}
}

@article{king2008dominance,
  title={Dominance and affiliation mediate despotism in a social primate},
  author={King, Andrew J and Douglas, Caitlin MS and Huchard, Elise and Isaac, Nick JB and Cowlishaw, Guy},
  journal={Current Biology},
  volume={18},
  number={23},
  pages={1833--1838},
  year={2008},
  publisher={Elsevier}
}

@article{nagy2010hierarchical,
  title={Hierarchical group dynamics in pigeon flocks},
  author={Nagy, M{\'a}t{\'e} and {\'A}kos, Zsuzsa and Biro, Dora and Vicsek, Tam{\'a}s},
  journal={Nature},
  volume={464},
  number={7290},
  pages={890--893},
  year={2010},
  publisher={Nature Publishing Group UK London}
}

@article{pinkoviezky2018collective,
  title={Collective conflict resolution in groups on the move},
  author={Pinkoviezky, Itai and Couzin, Iain D and Gov, Nir S},
  journal={Physical Review E},
  volume={97},
  number={3},
  pages={032304},
  year={2018},
  publisher={APS}
}

@article{attanasi2015emergence,
  title={Emergence of collective changes in travel direction of starling flocks from individual birds' fluctuations},
  author={Attanasi, Alessandro and Cavagna, Andrea and Del Castello, Lorenzo and Giardina, Irene and Jelic, Asja and Melillo, Stefania and Parisi, Leonardo and Pohl, Oliver and Shen, Edward and Viale, Massimiliano},
  journal={Journal of The Royal Society Interface},
  volume={12},
  number={108},
  pages={20150319},
  year={2015},
  publisher={The Royal Society}
}

@article{cavagna2010scale,
  title={Scale-free correlations in starling flocks},
  author={Cavagna, Andrea and Cimarelli, Alessio and Giardina, Irene and Parisi, Giorgio and Santagati, Raffaele and Stefanini, Fabio and Viale, Massimiliano},
  journal={Proceedings of the National Academy of Sciences},
  volume={107},
  number={26},
  pages={11865--11870},
  year={2010},
  publisher={National Acad Sciences}
}

@article{cavagna2013diffusion,
  title={Diffusion of individual birds in starling flocks},
  author={Cavagna, Andrea and Queir{\'o}s, SM Duarte and Giardina, Irene and Stefanini, Fabio and Viale, Massimiliano},
  journal={Proceedings of the Royal Society B: Biological Sciences},
  volume={280},
  number={1756},
  pages={20122484},
  year={2013},
  publisher={The Royal Society}
}

@article{wissner2013causal,
  title={Causal entropic forces},
  author={Wissner-Gross, Alexander D and Freer, Cameron E},
  journal={Physical review letters},
  volume={110},
  number={16},
  pages={168702},
  year={2013},
  publisher={APS}
}

@mastersthesis{Ebmeier2017path,
    author = {Ebmeier, F.},
    title =  {Simulations of pedestrian flow and lane formation through causal entropy},
    school = {Georg-August-Universität},
    year = {2017}
}

@article{stanley1968dependence,
  title={Dependence of critical properties on dimensionality of spins},
  author={Stanley, H Eugene},
  journal={Physical Review Letters},
  volume={20},
  number={12},
  pages={589},
  year={1968},
  publisher={APS}
}

@article{sampaio2024multidimensional,
  title={Multidimensional social influence drives leadership and composition-dependent success in octopus--fish hunting groups},
  author={Sampaio, Eduardo and Sridhar, Vivek H and Francisco, Fritz A and Nagy, M{\'a}t{\'e} and Sacchi, Ada and Strandburg-Peshkin, Ariana and N{\"u}hrenberg, Paul and Rosa, Rui and Couzin, Iain D and Gingins, Simon},
  journal={Nature Ecology \& Evolution},
  pages={1--13},
  year={2024},
  publisher={Nature Publishing Group UK London}
}

@article{strandburg2018inferring,
  title={Inferring influence and leadership in moving animal groups},
  author={Strandburg-Peshkin, Ariana and Papageorgiou, Danai and Crofoot, Margaret C and Farine, Damien R},
  journal={Philosophical Transactions of the Royal Society B: Biological Sciences},
  volume={373},
  number={1746},
  pages={20170006},
  year={2018},
  publisher={The Royal Society}
}

@article{peruani2017hydrodynamic,
  title={Hydrodynamic equations for flocking models without velocity alignment},
  author={Peruani, Fernando},
  journal={Journal of the Physical Society of Japan},
  volume={86},
  number={10},
  pages={101010},
  year={2017},
  publisher={The Physical Society of Japan}
}

@article{cucker2007emergent,
  title={Emergent behavior in flocks},
  author={Cucker, Felipe and Smale, Steve},
  journal={IEEE Transactions on automatic control},
  volume={52},
  number={5},
  pages={852--862},
  year={2007},
  publisher={IEEE}
}

@article{sumpter2008consensus,
  title={Consensus decision making by fish},
  author={Sumpter, David JT and Krause, Jens and James, Richard and Couzin, Iain D and Ward, Ashley JW},
  journal={Current Biology},
  volume={18},
  number={22},
  pages={1773--1777},
  year={2008},
  publisher={Elsevier}
}

\end{document}